\documentclass[intlimits,twoside,a4paper]{article}

\usepackage[cp1251]{inputenc}
\usepackage[eqsecnum]{cmpj3}

\issue{2025}{28}{1}{13601}
\doinumber{10.5488/CMP.28.13601}

\title[Cahn--Hilliard model with Schl\"ogl reactions. II.]%
{Cahn--Hilliard model with Schl\"ogl reactions: interplay of equilibrium and non-equilibrium phase
transitions. II. Memory effects}

\author[P. O. Mchedlov-Petrosyan,  L. N. Davydov]{P. O. Mchedlov-Petrosyan\orcid{0000-0002-2362-9077}\thanks{Corresponding author: \email{peter.mchedlov@free.fr}.}, L. N. Davydov\orcid{0000-0002-0031-2536}}

\address{A. I. Akhiezer Institute for Theoretical Physics, National Science Center ``Kharkiv Institite of
Physics \& Technology'', 1 Akademicheskaya Str., 61108 Kharkiv, Ukraine  }

\date{Received September 12, 2024, in final form December 05, 2024}

\begin{document}

\maketitle

\begin{abstract}
The present work is a continuation of our previous paper [Condens. Matter Phys., 2020, \textbf{23}, 33602: 1--17]. It is devoted to the modelling of the interplay of equilibrium and non-equilibrium phase transitions. The modelling of equilibrium phase transition is based on the modified Cahn--Hilliard equation. The non-equilibrium phase transition is modeled by the Second Schl\"ogl reaction system. We consider the advancing front, which combines these both transitions. Different from the first article, we consider here the memory effects, i.e., the effects of non-Fickian diffusion. The traveling wave solution is obtained, and its dependence on the model parameters is studied in detail. The relative importance of memory effects for different process regimes is estimated.

\end{abstract}

\section{Introduction}\label{Sec:1}

The present work is a continuation of our paper \cite{1}. It was devoted to the modelling of the interplay of equilibrium and non-equilibrium phase transitions. We considered the advancing fronts which ``combine'', in some sense, these both transitions. The equilibrium phase transition was modeled on the basis of convective-viscous Cahn--Hilliard equation \cite{2,3}. The non-equilibrium phase transition was modeled by canonical Schl\"ogl chemical reaction systems \cite{4}. The insight into the history and existing modifications of the Cahn--Hilliard equation was given in \cite{1} and will not be repeated here. We also refer to the pioneering papers and excellent reviews \cite{5,6,7,8}. Below, in the introduction, we first give only a  brief explanation of some basic assumptions and a description of the models, introduced in \cite{1}; then we outline the specificities of the present work.

As it was mentioned in \cite{1}, the basic underlying idea of Cahn--Hilliard model is that for an inhomogeneous system, e.g., system undergoing a phase transition, the thermodynamic potential (e.g., free energy) should depend not only on the order parameter $X$, but also on its gradient. For an inhomogeneous system, the local chemical potential $\mu $, defined as variational derivative of the thermodynamic potential functional, is as follows:
\begin{equation} \label{1.1} 
	\mu =-\bar{\varepsilon }^{2} \Delta X+f(X).  
\end{equation}
Herein our notations differ from notations in \cite{1}. $\bar{\varepsilon }$ is usually assumed to be proportional to the capillarity length, and $f\left(X\right)={\rd\Phi \left(X\right)}/{\rd X} $, where $\Phi \left(X\right)$ is a homogeneous part of the thermodynamic potential. As in \cite{1}, in the present communication we take $f\left(X\right)$ in the form of the cubic polynomial (corresponding to the fourth-order polynomial for a homogeneous part of thermodynamic potential):
\begin{equation} \label{1.2} 
	f\left(X\right)=qX^{3} -\bar{\delta }X^{2} -sX .   
\end{equation}
In this phenomenological model, we always consider the isothermal situation. So we do not show the temperature dependence of the coefficients in \eqref{1.2} explicitly. However, if we want to model the approach to a  critical state, for such a model the approach to critical temperature will be manifested by merging the stationary states together \cite{1}, i.e., by two non-zero roots of the right-hand side of \eqref{1.2} approaching the third zero root.

The diffusional flux $J$ is proportional to the gradient of chemical potential $\nabla \mu $; the proportionality coefficient is called mobility $M$~\cite{9}:
\begin{equation} \label{1.3} 
	J=-M\nabla \mu  .  
\end{equation}
Substitution of \eqref{1.3} into the continuity equation
\begin{equation} \label{1.4} 
	\frac{\partial X}{\partial t'} =-\nabla J 
\end{equation}
yields, with expression \eqref{1.1} for the chemical potential, instead of the usual second order diffusion equation, a fourth-order PDE for the order parameter $X$:
\begin{equation} \label{1.5} 
	\frac{\partial X}{\partial t'} =\nabla \left[M\nabla \mu \right] .
\end{equation}
In the presence of the external field, the convective term is added to equation \eqref{1.5}, and to account for dissipation the viscous term is added to the chemical potential $\mu $ \cite{10,11,12,13,2,3}:
\begin{equation} \label{1.6} 
	\frac{\partial X}{\partial t'} -\bar{\alpha }X\frac{\partial X}{\partial x'} =\frac{\partial }{\partial x'} \left[M\frac{\partial }{\partial x'} \left(\mu +\bar{\eta }\frac{\partial X}{\partial t'} \right)\right] .  
\end{equation}
Here, $\bar{\alpha }$ is proportional to the field and $\bar{\eta }$ is viscosity; we consider the one-dimensional problem herein.

Among numerous modifications of the Cahn--Hilliard (CH) equation we are interested here in the models, where the nonlinear sink/source terms, e.g., due to a chemical reaction, are inserted into the right-hand sides of \eqref{1.4} (and, correspondingly of \eqref{1.5} as well):
\begin{equation} \label{1.7} 
	\frac{\partial X}{\partial t'} =\nabla \left[M\nabla \mu \right]+R\left(X\right) .  
\end{equation}
For the classic CH equation such a study was pioneered by Huberman \cite{14}, Cohen and Murray \cite{15}; see also \cite{16,17,18,19,20,21}.

General observation is that the presence of chemical reaction can visibly influence the equilibrium phase transition, e.g., freeze the spinodal decomposition or coarsening, stabilizing some stationary inhomogeneous state. On the other hand, the canonical models for non-equilibrium phase transitions in chemical reaction systems were introduced by Schl\"ogl \cite{4}; here, the different ``phases'' correspond to different stationary states of the system. Schl\"ogl considered two reaction systems, the so-called ``First Schl\"ogl Reaction'' and the ``Second Schl\"ogl Reaction''. The first reaction exhibits a non-equilibrium phase transition of the second order, the second one exhibits a phase transition of the first order (for the details see \cite{4}).  If the system simultaneously undergoes an equilibrium phase transition accompanied by a phase separation, it could be of considerable interest to study the interaction of an equilibrium and non-equilibrium phase transitions. We also call such a process ``a combined transition'' for brevity.

In \cite{1} we considered the convective-viscous Cahn--Hilliard equation \eqref{1.6}, see \cite{2,3}, complemented by source/sink term $R\left(X\right)$. We considered $R\left(X\right)$ corresponding both to first and second Schl\"ogl reactions. We called these modifications Cahn--Hilliard--Huberman--Cohen--Murray (CHHCM) and Cahn--Hilliard--Schl\"ogl (CHS) equations, respectively. We obtained exact travelling-wave solutions for these modifications. For convective-viscous equation with the first Schl\"ogl reaction, the conditions of simultaneous equilibrium and non-equilibrium phase transitions are very restrictive and rigid. Even more, the equation for the travelling wave becomes degenerate in the absence of an applied field, splitting into second and first order equations. Thus, the constant-velocity combined-transition-front model is not very instructive in this case. On the other hand, for the CHS, the effect of the non-equilibrium transition, i.e., of the reactive system, is much stronger. The transition front may be stopped, or even reversed both by changing the stationary states of the reaction system and by the field. Thus, exploring the ``memory effects'', see below, in the present work we consider the convective-viscous CHS only.

The description of the ``Second Schl\"ogl Reaction'' is given in \cite{1}, equations~(1.10), (1.11). For the second Schl\"ogl reaction in the absence of diffusion, the evolution of $X$ is described by
\begin{equation} \label{1.8} 
	\frac{\rd X}{\rd t'} =\, -k'_{12} X^{3} +k_{12} AX^{2} -k_{22} BX+k'_{22} C .  
\end{equation}
We denote the concentrations with the same letters as species; the concentrations of species $A$, $B$ and $C$ (which are called the ``reservoir reagents'') are assumed to be constant and only concentration of $X$ can vary with time and space. We denote the rate constants by $k_{ij}$, $k'_{ij}$ for the forward and reverse reactions, respectively; to resemble with notations in \cite{1}, we keep the second lower index ``2'' for the second Schl\"ogl reaction. In \cite{1} we considered convective viscous CH equation with $R(X)$ equal to the right-hand-side of \eqref{1.8}.

However, in all models described above, the diffusional flux was assumed to be proportional to the gradient of the chemical potential. This ``modified'' Fick's law (the original Fick's law presumes proportionality to the gradient of concentration) was often criticized for the infinite speed of the spread of a diffusing substance. The most popular alternative is the Maxwell-Cattaneo approach, see \cite{22} for a comprehensive discussion; here, we give only the one-dimensional formulae, in accord with the spirit of the present paper. In Maxwell-Cattaneo approach, the mass-conservation, or continuity not changed. However, instead of \eqref{1.3} the following relation is proposed:
\begin{equation} \label{1.9} 
	\tau '\frac{\partial J}{\partial t'} +J=-M\frac{\partial \mu }{\partial x'}  .  
\end{equation}
Direct integration of the latter expression for the flux yields
\begin{equation} \label{1.10)} 
	J=-\int _{0}^{t'}\left(\frac{M}{\tau '} \frac{\partial \mu }{\partial x'} \right) \exp \left(\frac{t''-t'}{\tau '} \right)\rd t'' .  
\end{equation}
Thus, this approach is also called ``diffusion with memory effects''. Correspondingly, $\tau '$ is considered to be a characteristic time of the memory. On the other hand, eliminating $J$ from \eqref{1.9} yields the ``hyperbolic modification'' of the diffusion equation with a source
\begin{equation} \label{1.11} 
	\frac{\partial X}{\partial t'} +\tau '\frac{\partial }{\partial t'} \left[\frac{\partial X}{\partial t'} -R\left(X\right)\right]=\frac{\partial }{\partial x'} \left[M\frac{\partial \mu }{\partial x'} \right]+R\left(X\right).  
\end{equation}
Similarly, a hyperbolic modification of the convective-viscous Cahn--Hilliard equation \eqref{1.6} with source/sink term takes the form
\begin{equation} \label{1.12} 
	\frac{\partial X}{\partial t'} -\bar{\alpha }X\frac{\partial X}{\partial x'} +\tau '\frac{\partial }{\partial t'} \left[\frac{\partial X}{\partial t'} -R\left(X\right)\right]=\frac{\partial }{\partial x'} \left[M\frac{\partial }{\partial x'} \left(\mu +\bar{\eta }\frac{\partial X}{\partial t'} \right)\right]+R\left(X\right).   
\end{equation}
The hyperbolic modification of the classical Cahn--Hilliard equation was proposed in \cite{23} to model a rapid spinodal decomposition in a binary alloy. However, from purely mathematical point of view --- as a singular perturbation of the classic Cahn--Hilliard equation --- it was considered even earlier \cite{24}. These papers were followed by many others, both of physical and mathematical nature \cite{25,26,27,28,29}. For hyperbolic convective-viscous Cahn--Hilliard equation without source/sink term, i.e., \eqref{1.12} with $R(X)=0$, the exact travelling-wave solution was given in \cite{3}.

This paper is organized as follows: in the next section we give an exact travelling wave solution for the convective-viscous CHS equation complemented by memory effects. In section~\ref{Sec:3} we study the parametric dependence of this solution. In section~\ref{Sec:4} we discuss our results.

\section{Convective viscous Cahn--Hilliard--Schl\"ogl equation with memory effects}\label{Sec:2}

In the present section we first give exact travelling-wave solutions for convective viscous Cahn--Hilliard equation \cite{2,3} supplemented by memory effects and third order reaction terms. Thus, we first take into account the influence of both external field and dissipation; then we drop the convective and viscous terms setting $\bar{\alpha }=0$, $\bar{\eta }=0$, thereby reducing the equation to hyperbolic Cahn--Hilliard--Schl\"ogl equation.

To make the model and calculations somewhat more transparent we assume the  reaction to be irreversible, i.e., $k'_{22} =0$ in \eqref{1.8}. Substituting the right-hand side of \eqref{1.8} into \eqref{1.12} for $R\left(X\right)$,  we write down the convective viscous CHS equation with the memory effect, first in terms of the initial variable~$X$ (concentration):
\begin{eqnarray}  \label{2.1} 
	& &{\frac{\partial X}{\partial t'} -\bar{\alpha }X\frac{\partial X}{\partial x'} +\tau '\frac{\partial }{\partial t'} \left[\frac{\partial X}{\partial t'} -\left(-k'_{12} X^{3} +k_{12} AX^{2} -k_{22} BX\right)\right]} \nonumber \\
 & &=M\frac{\partial ^{2} }{\partial x'^{2} } \left(\mu +\bar{\eta }\frac{\partial X}{\partial t'} \right)\, -k'_{12} X^{3} +k_{12} AX^{2} -k_{22} BX, 
 \end{eqnarray}
\begin{equation} \label{2.2} 
	\mu =-\bar{\varepsilon }^{2} \frac{\partial ^{2} X}{\partial x'^{2} } +qX^{3} -\bar{\delta }X^{2} -sX.   
\end{equation}
Writing down equations \eqref{2.1}--\eqref{2.2} we implicitly assume that in the system $A-B-C-X$ the components~$A$ and $B$ are in large excess and are not essentially exhausted during the chemical reaction; we also presumed~$M$ to be a constant. Renormalizing $X$, $x'$ and $t'$, we introduce

\begin{equation} \label{2.3)} 
	X=uX_{0} ;\quad x'=xL; \quad t'=tT.   
\end{equation}
Here, $X_{0} ={1}/{\sqrt{q} } $, $T={1}/{k'_{12} X_{0}^{2} } ={q}/{k'_{12} } $ and $L=\sqrt{MT} =\sqrt{{M}/{k'_{12} X_{0}^{2} } } =\sqrt{{Mq}/{k'_{12} } } $.   Denoting $\alpha =\bar{\alpha }({X_{0} T}/{L}) =\bar{\alpha }({1}/{\sqrt{k'_{12} M} })$; $\varepsilon ^{2} ={\bar{\varepsilon }^{2} }/{L^{2} } $; $\eta ={\bar{\eta }}/{T} $; $\delta =X_{0} \bar{\delta }= {\bar{\delta }}/{\sqrt{q} }$;    $\Theta ={k_{12} A}/{k'_{12} X_{0} } $, $\Omega ={k_{22} B}/{k'_{12} X_{0}^{2} } $ and $\tau ={\tau '}/{T} $, we write down equation \eqref{2.1} in non-dimensional form
\begin{eqnarray} \label{2.4} 
	& &{\frac{\partial u}{\partial t} -\alpha u\frac{\partial u}{\partial x} +\tau \frac{\partial }{\partial t} \left[\frac{\partial u}{\partial t} +u\left(u^{2} -\Theta u+\Omega \right)\right]} \nonumber \\
	& &=\frac{\partial ^{2} }{\partial x^{2} } \left(-\varepsilon ^{2} \frac{\partial ^{2} u}{\partial x^{2} } +u^{3} -\delta u^{2} -su+\eta \frac{\partial u}{\partial t} \right)-u\left(u^{2} -\Theta u+\Omega \right).  
\end{eqnarray}
Below we assume that the quadratic equation
\begin{equation} \label{2.5)} 
	u^{2} -\Theta u+\Omega =0 
\end{equation}
always has real roots $u_{1}$, $u_{2}$, $0<u_{1} \leqslant u_{2} $, corresponding to stationary states of the reactions system. I.e., $\Theta ^{2} -4\Omega \geqslant 0$ which means, in terms of the parameters of the reaction system, $\left(k_{12} A\right)^{2} \geqslant 4k'_{12} k_{22} B$. This is simply the necessary condition of the multiplicity of stationary states in Schl\"ogl reaction model, see equations (1.10), (1.11) of \cite{1}, and equation \eqref{1.8}, i.e., the existence of a non-equilibrium phase transition.

Looking for the traveling wave solutions of \eqref{2.4}, we introduce the travelling wave coordinate $z=x-vt$. This yields
\begin{eqnarray} \label{2.6}  
	& &\frac{\rd}{\rd z} \left[vu+\alpha \frac{u^{2} }{2} +v\tau u\left(u-u_{1} \right)\left(u-u_{2} \right)\right.  \nonumber \\
	& &{\left. +\frac{\rd}{\rd z} \left(-v^{2} \tau u-\varepsilon ^{2} \frac{\rd^{2} u}{\rd z^{2} } +u^{3} -\delta u^{2} -su-v\eta \frac{\rd u}{\rd z} \right)\right]=u\left(u-u_{1} \right)\left(u-u_{2} \right)}.
\end{eqnarray}
We look for the solution, which connects the stable stationary state of the reaction system $u=u_{2} $ at $z=-\infty $ with the stable stationary state $u=0$ at $z=+\infty $. Thus, the proper \textit{Ansatz} for the anti-kink solution (as usually we call ``kinks'' the solutions with ${\rd u}/{\rd z} >0$, and ``anti-kinks'' --- the solutions with ${\rd u}/{\rd z} <0$) having this property will be
\begin{equation} \label{2.7} 
	\frac{1}{\kappa } \frac{\rd u}{\rd z} =u\left(u-u_{2} \right),   
\end{equation}
where $\kappa $ is presently unknown positive constant. Assuming that the solutions of \eqref{2.7} are simultaneous solutions of equation \eqref{2.6}, we can rewrite \eqref{2.6} as follows:
\begin{eqnarray} \label{2.8)}   
	& & {\frac{\rd}{\rd z} \left[vu+\alpha \frac{u^{2} }{2} \right. -\left(\frac{1}{2\kappa } u^{2} -\frac{u_{1} }{\kappa } u\right)} \nonumber \\
	& &{\left. +\frac{\rd}{\rd z} \left(\frac{v\tau }{2\kappa } u^{2} -\frac{u_{1} v\tau }{\kappa } u-\varepsilon ^{2} \frac{\rd^{2} u}{\rd z^{2} } +u^{3} -\delta u^{2} -\left(v^{2} \tau +s\right)u-v\eta \frac{\rd u}{\rd z} \right)\right]=0}.  
\end{eqnarray}
Integrating once, we get
\begin{eqnarray} \label{2.9} 
	& & {\left(v+\frac{u_{1} }{\kappa } \right)u+\left(\alpha -\frac{1}{\kappa } \right)\frac{u^{2} }{2} } \nonumber \\
	& &{+\frac{\rd}{\rd z} \left[-\varepsilon ^{2} \frac{\rd^{2} u}{\rd z^{2} } +u^{3} +\left(\frac{v\tau }{2\kappa } -\delta -v\eta \kappa \right)u^{2} -\left(v^{2} \tau +s+\frac{u_{1} v\tau }{\kappa } -v\eta \kappa u_{2} \right)u\right]=C_{1} }, 
\end{eqnarray}
where $C_{1} $ is an arbitrary constant. The expression for the second derivative of $u$ is given by
\begin{equation} \label{2.10)} 
	\frac{\rd^{2} u}{\rd z^{2} } =\kappa ^{2} \left(2u^{3} -3u_{2} u^{2} +u_{2}^{2} u\right). 
\end{equation}
Regarding the \textit{Ansatz} \eqref{2.7}, for the equation \eqref{2.9} to be satisfied, the expression under the derivative should be linear in $u$. I.e., for \eqref{2.7}, to get a solution of \eqref{2.6}, two following equations should be \textit{identically} satisfied for \textit{arbitrary} $u$:
\begin{equation} \label{2.11} 
	\left(v+\frac{u_{1} }{\kappa } -\beta \kappa u_{2} \right)u+\left(\alpha -\frac{1}{\kappa } +2\beta \kappa \right)\frac{u^{2} }{2} =C_{1} ,
\end{equation}
\begin{eqnarray} \label{2.12)} 
	& &{\left(1-2\varepsilon ^{2} \kappa ^{2} \right)u^{3} +\left(3\varepsilon ^{2} \kappa ^{2} u_{2} +\frac{v\tau }{2\kappa } -\delta -v\eta \kappa \right)u^{2} } \nonumber \\
	& &{-\left(\varepsilon ^{2} \kappa ^{2} u_{2}^{2} +v^{2} \tau +s+\frac{u_{1} v\tau }{\kappa } -v\eta \kappa u_{2} +\beta \right)u=C_{2} },  
\end{eqnarray}
where $\beta $ is an unknown constant. Equating to zero coefficients at each power of $u$, including $C_{1}$, $C_{2} $ as zero power coefficients, we finally obtain five constraints on the parameters:
\begin{equation} \label{2.13} 2\varepsilon ^{2} \kappa ^{2} =1,   \end{equation}
\begin{equation} \label{2.14} \alpha -\frac{1}{\kappa } +2\beta \kappa =0 ,  \end{equation}
\begin{equation} \label{2.15} 3\varepsilon ^{2} \kappa ^{2} u_{2} +\frac{v\tau }{2\kappa } -\delta -v\eta \kappa =0,   \end{equation}
\begin{equation} \label{2.16} v+\frac{u_{1} }{\kappa } -\beta \kappa u_{2} =0,   \end{equation}
\begin{equation} \label{2.17} \varepsilon ^{2} \kappa ^{2} u_{2}^{2} +v^{2} \tau +s+\frac{u_{1} v\tau }{\kappa } -v\eta \kappa u_{2} +\beta =0.   \end{equation}
If the constraints \eqref{2.13}--\eqref{2.17} are fulfilled, the solution of \eqref{2.7} is simultaneously solution of \eqref{2.1}. Integrating \eqref{2.7} and taking $z=0$ for maximal steepness point, we get
\begin{equation} \label{2.18} 
	u=\frac{u_{2} }{2} \left\{1-\tanh\left[\frac{u_{2} }{2\sqrt{2} \varepsilon } \left(x-vt\right)\right]\right\}. 
\end{equation}
The functional form of a solution is rather simple; however, the dependence of the parameters of solution on the system parameters is quite complicated, see~\cite{1}. In the next section we, following the lines of~\cite{1}, consider the changes introduced by the memory.

\section{The parametric dependence of solution}\label{Sec:3}

Similarly to \cite{1} there are five constraints, and only three unknowns $\kappa ,\, v$ and $\beta $. I.e., for the constant velocity transition front to exist, two additional constraints on the values of the stationary states of the reaction system and on the values of the equilibrium states for the phase transition should be imposed. We assume, as in \cite{1}, that the parameters related to the reaction system are ``basic''. Evidently, \eqref{2.13}--\eqref{2.14} and \eqref{2.16}, and correspondently $\kappa$,  $v$, $\beta$ coincide with the given in Part~I~\cite{1} (we remind that notations are different):
\begin{equation} \label{3.1)} 
	\kappa ^{2} =\frac{1}{2\varepsilon ^{2} } , 
\end{equation}
\begin{equation} \label{3.2)} 
	\beta =\frac{1}{2\kappa ^{2} } -\frac{\alpha }{2\kappa },  
\end{equation}
\begin{equation} \label{3.3} 
	v=\frac{1}{\kappa } \left(\frac{u_{2} }{2} -u_{1} \right)-\frac{\alpha u_{2} }{2} .   
\end{equation}
Naturally, for $\tau =0$, the equations \eqref{2.15}, \eqref{2.17} coincide with the corresponding equations of Part I \cite{1} too. We denote the values, corresponding to $\tau =0$ as $\delta _{0}$ and $s_{0} $:

\begin{equation} \label{3.4} 
	\delta _{0} =\, \, \frac{3}{2} u_{2} -\left[\left(\frac{u_{2} }{2} -u_{1} \right)-\kappa \frac{\alpha u_{2} }{2} \right]\eta,
\end{equation}
\begin{equation} \label{3.5} 
	s_{0} =\eta u_{2} \left(\frac{u_{2} }{2} -u_{1} \right)-\frac{1}{2} \left(u_{2}^{2} +\, \frac{1}{\kappa ^{2} } \right)+\frac{\alpha }{2} \left(\frac{1}{\kappa } -\eta \kappa u_{2}^{2} \right).   
\end{equation}
We look for solutions of \eqref{2.15}, \eqref{2.17} in the form
\begin{equation} \label{3.6)} 
	\delta =\delta _{0} +\tau \delta _{1}, \quad s=s_{0} +\tau s_{1} .   
\end{equation}
This yields

\begin{equation} \label{3.7}  
	\delta _{1} =\frac{v}{2\kappa } ,   
\end{equation}
\begin{equation} \label{3.8} 
	s_{1} =-v^{2} -\frac{u_{1} v}{\kappa } =-v\left(v+\frac{u_{1} }{\kappa } \right).   
\end{equation}
Naturally, the corrections due to the memory are proportional to the velocity of the front; there should be no corrections for a static front.

The constraints \eqref{2.15} and \eqref{2.17} impose some limitations on the parameters $\delta $ and $s$ of the equilibrium phase transition, see \eqref{1.2}, i.e., on the roots $\tilde{u}_{1}$, $\tilde{u}_{2}$,  $\tilde{u}_{3} $ of the cubic equation
\begin{equation} \label{3.9} 
	\tilde{u}\left(\tilde{u}^{2} -\delta \tilde{u}-s\right)=0.   
\end{equation}
These roots are positions of the extrema of the homogeneous part of the thermodynamic potential $\Phi \left(u\right)$; we denote the roots, corresponding to stable minima $\, \tilde{u}_{2}$, $\tilde{u}_{3} $, and $\tilde{u}_{1} $ --- corresponding to an unstable maximum. The root $\tilde{u}_{3} =0$ coincides with one of the stationary states of the reaction system. The expressions for two remaining roots $\tilde{u}_{1} $ and $\tilde{u}_{2} $ yield the constraints imposed on the stationary values for the equilibrium transformation:
\begin{equation} \label{3.10} 
	\tilde{u}_{2,1} =\frac{1}{2} \left[\left(\delta _{0} +\tau \delta _{1} \right)\pm \sqrt{\left(\delta _{0} +\tau \delta _{1} \right)^{2} +4\left(s_{0} +\tau s_{1} \right)} \right].   
\end{equation}
We are mainly interested in what extent the actual final concentration $u_{2} $ after the transition deviates from the equilibrium value $\, \tilde{u}_{2} $. Usually $\tau $ is assumed to be a small parameter; expanding \eqref{3.10} and keeping only linear in $\tau $ terms, we have got
\begin{equation} \label{3.11} 
	\tilde{u}_{2} =\frac{1}{2} \left[\delta _{0} +\sqrt{\delta _{0}^{2} +4s_{0} } +\left(\delta _{1} +\frac{\delta _{0} \delta _{1} +2s_{1} }{\sqrt{\delta _{0} ^{2} +4s_{0} } } \right)\tau \right] .  
\end{equation}
Then, the deviation of the equilibrium value from the actual one after the transition is
\begin{equation} \label{3.12)} 
	\tilde{u}_{2} -u_{2} =\frac{1}{2} \left[\left(\delta _{0} +\sqrt{\delta _{0}^{2} +4s_{0} } -2u_{2} \right)+\left(\delta _{1} +\frac{\delta _{0} \delta _{1} +2s_{1} }{\sqrt{\delta _{0} ^{2} +4s_{0} } } \right)\tau \right] .  
\end{equation}
The $\tau $-independent term was calculated and discussed in \cite{1}; here, we are interested in the relative importance of the memory effect
\begin{eqnarray} \label{3.13)} 
	& & {\tilde{u}_{2} -u_{2} =\frac{1}{2} \left(\delta _{0} +\sqrt{\delta _{0}^{2} +4s_{0} } -2u_{2} \right) } \nonumber \\
	& &{\times \left[1+\left(\delta _{0} +\sqrt{\delta _{0}^{2} +4s_{0} } -2u_{2} \right)^{-1}  \left(\delta _{1} +\frac{\delta _{0} \delta _{1} +2s_{1} }{\sqrt{\delta _{0} ^{2} +4s_{0} } } \right)\tau \right]}.  
\end{eqnarray}
Thus, we define
\begin{equation} \label{3.14} 
	F=\left(\delta _{0} +\sqrt{\delta _{0}^{2} +4s_{0} } -2u_{2} \right)^{-1} \left(\delta _{1} +\frac{\delta _{0} \delta _{1} +2s_{1} }{\sqrt{\delta _{0} ^{2} +4s_{0} } } \right).   
\end{equation}
We study the behavior of $F$ as a function of the system parameters. Following \cite{1}, we consider several special cases. First we consider CHS, i.e., \eqref{2.4}, where the applied field and dissipation are absent, $\alpha =0;\, \, \eta =0$. Then, for $\tau =0$ we have
\begin{eqnarray} \label{3.15} 
	& & {\left. \left(\tilde{u}_{2} -u_{2} \right)\right|_{\tau =0} =\frac{1}{2} \left(\delta _{0} +\sqrt{\delta _{0}^{2} +4s_{0} } -2u_{2} \right)} \nonumber \\
	 & &{=\frac{1}{2} \left(\frac{3}{2} u_{2} +\, \frac{1}{2} u_{2} \sqrt{1-\frac{8}{u_{2}^{2} \kappa ^{2} } } -2u_{2} \right)=\frac{u_{2} }{4} \left( \sqrt{1-\frac{8}{u_{2}^{2} \kappa ^{2} } } -1\right)}.  
 \end{eqnarray}
From \eqref{3.15} an evident limitation arises (see \cite{1}): ${8}/{u_{2}^{2} \kappa ^{2} } <1;$ or $\varepsilon <{u_{2} }/{4} $. It is, however, not a severe limitation, it is usually assumed $\varepsilon \ll 1$. Calculating $F$ given by \eqref{3.14}, we get
\begin{equation} \label{3.16} 
	F=\varepsilon ^{2} \frac{1-2\frac{u_{1} }{u_{2} } }{\sqrt{1-16\frac{\varepsilon ^{2} }{u_{2}^{2} } } }  .  
\end{equation}
I.e., the correction due to the memory has always the same sign as velocity (for positive $\kappa $); its relative importance increases with velocity and $\varepsilon $ (increasing $\varepsilon $ means decreasing steepness of the front).

Now, let us consider the convective CHS equation: the field is non-zero, $\alpha \ne 0;$ but the viscosity is still zero, $\eta =0$. It is worth noting that there is an upper limit on $\alpha $ in this case, corresponding to $s_{0} =0$:
\begin{equation} \label{3.17} 
	\alpha _{l} =\kappa u_{2}^{2} \left(1+\frac{1}{\kappa ^{2} u_{2}^{2} } \right) .  
\end{equation}
If $\alpha \to \alpha _{l} $, the $\tilde{u}_{1} $ root of \eqref{3.9} (unstable maximum of the potential) approaches zero (stable minimum); it means merging of stable and unstable states in the thermodynamic potential. Thus, the combined transition front is impossible for $\alpha >\alpha _{l} $.

\begin{figure}
	\begin{center}
\includegraphics[width=0.6\textwidth]{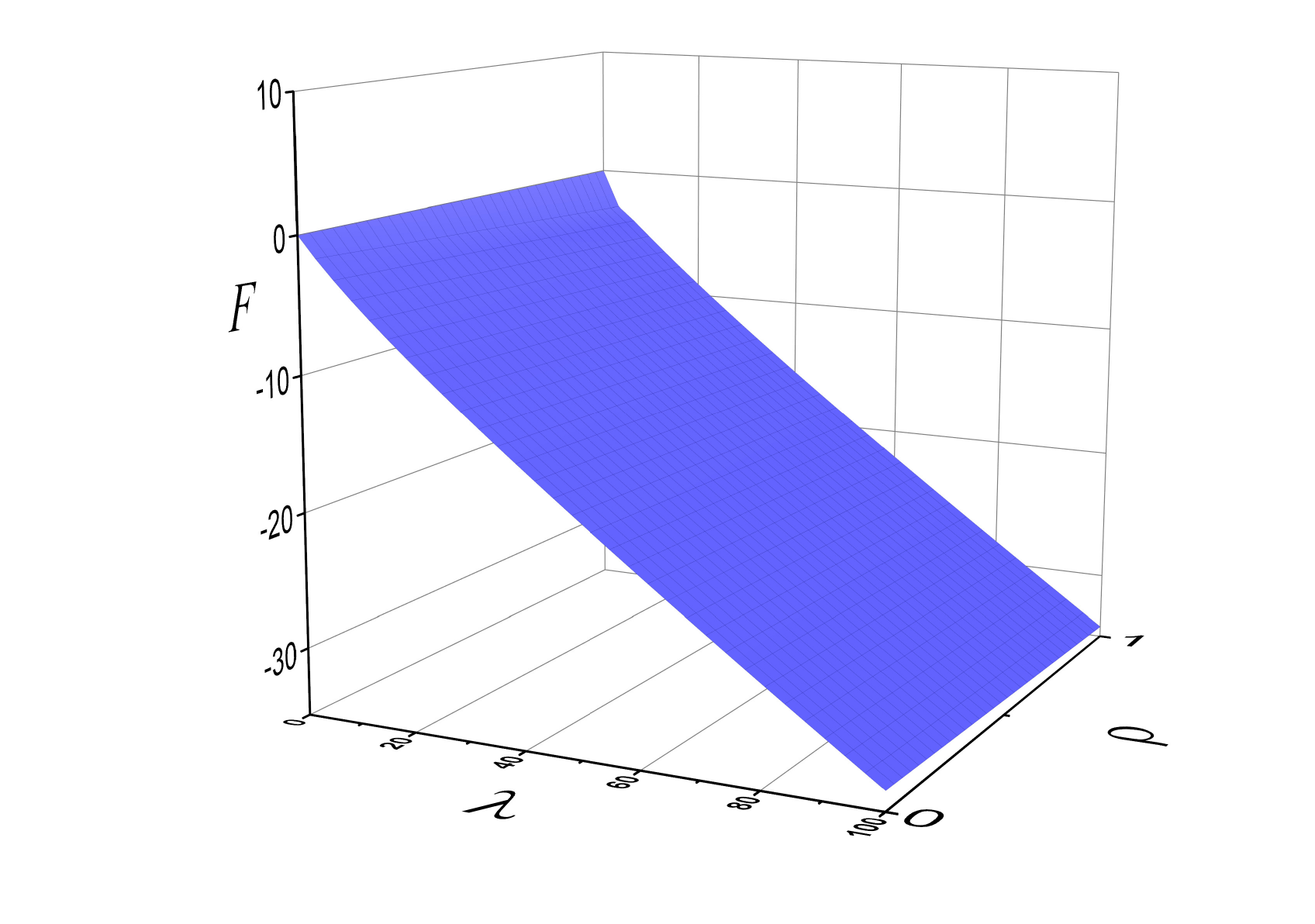}
\caption{(Colour online) Dependence of $F$ on parameters $\lambda = \alpha \kappa$ and $\rho = u_1/u_2 $ according to \eqref{3.18}. The values of $\kappa$ and $u_2$ are taken by way of example as $\kappa =10$ and $u_2=1$.  }
\label{Fig:1}
\end{center}
\end{figure}

According to \eqref{3.14}
\begin{equation} \label{3.18} 
	F=\frac{\left(1-2\frac{u_{1} }{u_{2} } \right)-\alpha \kappa }{2\kappa ^{2} \left[\sqrt{1+8\left(\frac{\alpha }{\kappa u_{2}^{2} } -\frac{1}{\kappa ^{2} u_{2}^{2} } \right)} -1\right]} \left[1+\frac{4\alpha \kappa -1}{\sqrt{1+8\left(\frac{\alpha }{\kappa u_{2}^{2} } -\frac{1}{\kappa ^{2} u_{2}^{2} } \right)} } \right].
\end{equation}
Numeric example for \eqref{3.18} is given in figure \ref{Fig:1}. The equation \eqref{3.18} could be simplified further.  First, $\alpha \gg \frac{1}{\kappa } \sim \varepsilon \ll 1 $. Then, according to \eqref{3.17} always $\alpha <\alpha _{l} $; so we can presume that $\alpha $ is not too close to its upper limit (which corresponds to extremely asymmetric potential), $\alpha \ll \alpha_{l} $, and $8({\alpha }/{\kappa u_{2}^{2} }) =8\sqrt{2} ({\alpha \varepsilon }/{u_{2}^{2} }) \ll 1$. Thus, equation \eqref{3.18} is approximately
\begin{equation} \label{3.19} 
	F\simeq \frac{2}{\kappa ^{2} }\left[\alpha \kappa -\left(1-2\frac{u_{1} }{u_{2} } \right)\right]\left(\alpha \kappa -\frac{\kappa ^{2} u_{2}^{2} +1}{4} \right).  
\end{equation}

According to the latter expression, the relative correction due to the memory seems to be non-monotonous, has a minimum, and changes the sign at $\alpha \kappa = y_1, y_2$:
\begin{equation} \label{3.20)} 
	y_{1} =\left(1-2\frac{u_{1} }{u_{2} } \right); \quad y_{2} =\frac{\kappa ^{2} u_{2}^{2} +1}{4} .   
\end{equation}
Per definition $-1<1-2\frac{u_{1} }{u_{2} } <1$; so $\alpha \kappa $ can transverse $y_1$; it also means the change of the velocity sign. However, according to our assumption, ${\alpha }/{\kappa u_{2}^{2} } \ll 1;$ i.e. $\alpha \kappa \ll \frac{1}{2} \left( y_1+y_2 \right)\simeq \frac{1}{8} \kappa ^{2} u_{2}^{2} $. That is, when $\alpha \kappa $ increases, the relative correction due to the memory will be positive and will have a decreasing absolute value for $\alpha \kappa < y_1 $, and will be negative with an increasing absolute value for $y_{1} <\alpha \kappa $. I.e., the dependence is monotonous in the proper domain, see figure \ref{Fig:1}, where $\alpha \kappa$  will not reach the position of minimum of~\eqref{3.19}.

Now, let us consider the viscous CHS equation: the field is zero, $\alpha =0$ but the viscosity is non-zero, $\eta \ne 0$. Then, from \eqref{3.4}--\eqref{3.5}
\begin{equation} \label{3.21} \delta _{0} =\, \frac{3}{2} u_{2} -\left(\frac{u_{2} }{2} -u_{1} \right)\eta ,   \end{equation}
\begin{equation} \label{3.22} s_{0} =\eta u_{2} \left(\frac{u_{2} }{2} -u_{1} \right)-\frac{1}{2} \left(u_{2}^{2} +\, \frac{1}{\kappa ^{2} } \right) .  \end{equation}
Similar to the case $\alpha \ne 0$, there is an upper limit on $\eta $ in this case, corresponding to $s_{0} =0$:
\begin{equation} \label{3.23} \eta _{l} =\frac{u_{2} +\, \frac{1}{\kappa ^{2} u_{2} } }{u_{2} -2u_{1} } .   \end{equation}
If $\eta \to \eta _{l} $, the $\tilde{u}_{1} $ root of \eqref{3.9} approaches zero; it means again the merging of stable and unstable states in the thermodynamic potential. Thus, the combined transition front is impossible for $\eta >\eta _{l} $.

Using \eqref{3.21} and \eqref{3.22}, we get
\begin{equation} \label{3.24} \delta _{0}^{2} +4s_{0} =\left(\frac{u_{2} }{2} +v\kappa \eta \right)^{2} -\frac{2}{\kappa ^{2} }  .  \end{equation}
Velocity \eqref{3.3} is not dependent on $\eta $, so the expressions \eqref{3.7} and \eqref{3.8} for $\delta _{1} $ and $s_{1} $ are the same as in the case $\alpha =0$; $\eta =0$. Then,
\begin{equation} \label{3.25} \delta _{0} \delta _{1} +2s_{1} =-\left(\frac{1}{2} u_{2} +v\kappa \eta \right)\frac{v}{2\kappa }  .  \end{equation}
Substitution of the expressions \eqref{3.24} and \eqref{3.25} into \eqref{3.14} yields
\begin{equation} \label{3.26} F=\frac{1}{2\kappa } \frac{v}{\sqrt{\left(\frac{u_{2} }{2} +v\kappa \eta \right)^{2} -\frac{2}{\kappa ^{2} } } }  .  \end{equation}
I.e., for positive $v$ the effect of the memory decreases monotonously with an increasing viscosity.

\section{Discussion}\label{Sec:4}

In the present work we studied the effect of the memory on the model, introduced in \cite{1}. It is a model of interplay of equilibrium and non-equilibrium phase transitions. Such an interplay was analyzed by considering the advancing front which ``combines'', in some sense, these both transitions. While the equilibrium phase transition was described by the modified \cite{2,3} Cahn--Hilliard equation, the non-equilibrium phase transition was presented by the canonical chemical models introduced by Schl\"ogl~\cite{4}.

In \cite{1} we considered the convective-viscous Cahn--Hilliard equation with additional nonlinear terms, corresponding both to the first and second Schl\"ogl reactions; there are quadratic and cubic nonlinearities, respectively. We obtained the exact travelling waves solutions for these equations.

The main idea was: for combined transition, i.e., for equilibrium and non-equilibrium transitions in order to proceed simultaneously, some constraints should be imposed on the parameters, linking the parameters of the ``Cahn--Hilliard part'' to the parameters of the chemical system. It was shown that the equilibrium stationary states of the thermodynamic potential in the ``Cahn--Hilliard part'' are necessarily connected to the stationary states of the chemical system. Generally they do not coincide; however, the actual stationary concentration after transition is that of the chemical system. Indeed, otherwise the equations could not be satisfied (the derivative terms disappear at $\pm \infty $). On the other hand, this means that with respect to the values of the equilibrium stationary states of the thermodynamic potential, the system will be over-, or undersaturated \cite{1}. I.e., the complete equilibration will be prevented.

As in \cite{1}, to make the model formulation possibly transparent, we assumed the last reaction of the system to be irreversible; indeed, by changing the kinetic parameter and/or diminishing the concentration~$C$, this reaction could always be shifted close to irreversible. On the other hand, for the reversible reaction, the equilibrium concentration \textit{before} transition will also deviate from the actual one; this will demand a special preparation of the system and make the model quite artificial.

In \cite{1} the description of the mass transfer was based on the ``generalized Fick's law''~\eqref{1.3}; in the present paper we introduce the memory effect, i.e., we use Maxwell-Cattaneo approach \eqref{1.9} \cite{23}. The most important result of the present work is that despite the introduction of the memory effects, the CHS equation still possesses an exact travelling wave solution. This is nontrivial: many well known diffusive model equations, e.g., Newell-Whitehead equation, Nagumo equation, Zeldovich equation etc., see \cite{30}, lose this property, when the memory effects are introduced, see \cite{31}.

The exact travelling wave solution is a ``tool'', which exhibits clearly the effect of the memory. The exact solution \eqref{2.18} has the same form, as in \cite{1}. Even more, the expressions for the parameters of solution are exactly the same. I.e., while for the non-stationary processes the memory effect can change the rates and the form of the wave, see, e.g., \cite{23}, for the constant velocity transition, the velocity and steepness of the wave are the same. However, the compatibility conditions \eqref{2.15}, \eqref{2.17} are changed. This changes correspondingly the possible over/undersaturation after the transition. The additional terms are proportional to the non-dimensional characteristic time $\tau $. In dimensional parameters, the expression for $\tau $ is
\begin{equation} \label{4.1)} 
	\tau =\frac{\tau '}{T} =\frac{\tau 'k'_{12} }{q}  .  
\end{equation}
Here, $\tau $ is the ratio of the initial characteristic time in Maxwell-Cattaneo equation to the characteristic time of the chemical reaction; it is usually considered to be a small parameter, $\tau \ll 1$. Thus, considering the expressions for the stationary states of thermodynamic potential \eqref{3.10}, we can always expand them in $\tau $ and keep the linear terms only, see \eqref{3.11}. We are interested in the changes, due to memory, of the deviations of the actual stationary concentration after transition from its equilibrium value. To measure these relative changes, we introduced the function $F$, see \eqref{3.14}. Of course, the relative change is $\tau F$, but we will drop $\tau $ herein.

Proceeding along the same lines as in \cite{1}, we considered three special cases. First we considered the CHS case when the applied field and the dissipation are absent, $\alpha =0;\, \, \eta =0$, i.e., the classic CH equation complemented by the cubic nonlinearity. The single limitation in this case is $4\varepsilon <u_{2} $, or
\begin{equation} \label{4.2)} 
	\frac{1}{4} u_{2} >\frac{\bar{\varepsilon }}{L} =\frac{\bar{\varepsilon }}{\sqrt{MT} } \ll 1.   
\end{equation}
Here, $\bar{\varepsilon }$ is of the order of capillarity length; $L$ is the diffusion distance for the characteristic time of the chemical reaction; so this limitation is not restrictive. From \eqref{3.16} it is evident, that the relative value of memory correction is zero for zero velocity of the front and is maximal for $u_{1} =0,$ that is when unstable state of the chemical system merges with the stable state $u=0$. It is also increasing with $\varepsilon $; increasing~$\varepsilon $ means decreasing steepness of the front. Indeed, for the fast moving shallow front, the relaxation to equilibrium is delayed, and the integrated over $\tau $ influence of the ``history'' increases.

If $\alpha \ne 0$; $\eta =0$, there is an upper limit on $\alpha $, see \eqref{3.17}, neglecting a small addition
\begin{equation} \label{4.3)} 
	\alpha _{l} \simeq \kappa u_{2}^{2} =u_{2}^{2} \frac{1}{\sqrt{2} \varepsilon } =u_{2}^{2} \frac{L}{\sqrt{2} \bar{\varepsilon }}  .  
\end{equation}
Thus, $\alpha _{l} $ is rather large; but if $\alpha \to \alpha _{l} $, the $\tilde{u}_{1} $ root of \eqref{3.9} approaches zero; it means merging of stable and unstable states in the thermodynamic potential. Thus, the combined transition front is impossible for $\alpha >\alpha _{l} $. The approximate expression \eqref{3.19} for $F$, as well as numerical example in figure~\ref{Fig:1}, show that with growing $\alpha \kappa $ the relative correction is positive and decreases for positive velocity, it is zero for zero velocity and it is negative with increasing absolute value for negative velocity. Thus, the memory effect became more pronounced, when the field forces the ``retreat'' of the front, acting against the normal direction determined by the chemical system. This again leads to a delayed relaxation.

Considering the viscous CHS equation, we have $\eta \ne 0$; $\alpha =0$. Similar to the case $\alpha \ne 0$ , there is an upper limit on $\eta $ in this case, \eqref{3.23}, corresponding to the merging of unstable $\tilde{u}_{1} $ and stable $\tilde{u}=0$ states of the thermodynamic potential. Neglecting small correction, this limit is as follows:
\begin{equation} \label{4.4)} 
	\eta _{l} \simeq \frac{u_{2} }{u_{2} -2u_{1} }  .  
\end{equation}
For the velocity approaching zero $\eta _{l} \to \infty $, although the exact expression \eqref{3.26} gives zero correction. For positive velocity, the correction decreases monotonously with increasing viscosity. For negative velocity, there is an additional limitation: if $\eta \to \eta _{m} $,
\begin{equation} \label{4.5)} 
	\eta _{m} =\frac{\kappa u_{2} -2\sqrt{2} }{2\kappa u_{1} -\kappa u_{2} }  ,  
\end{equation}
the unstable state $\tilde{u}_{1} $ of the potential merges with stable state $\tilde{u}_{2} $, see \eqref{3.9}, and the model becomes physically senseless. However, for $\eta <\eta _{m} $, the absolute value of the correction increases with viscosity~$\eta $, and for $\eta >\eta _{m} $, it  decreases.

Summing up, for positive velocity of the transition front, when the state with larger concentration advances, both the applied field and viscosity generally diminish the memory corrections. For negative velocity, the consistency of the model imposes stronger limitations on the allowed intervals of the parameters; the delayed relaxation makes the memory effects more pronounced. I.e., if the diffusion in the given system demonstrates the memory effects, changing the transition velocity, and the applied field yields an additional instrument to achieve the necessary over/undersaturation after transition.

\section*{Acknowledgements}

We are thankful to O. S. Bakai for the attention to the paper and for valuable remarks.

\newpage
\ukrainianpart

\title{Модель Кана--Хiльярда з реакціями  Шльогля: взаємодія рівноважного і нерівноважного
фазових переходів. ІI.~Ефекти пам'яті}
\author{П. О. Мчедлов-Петросян, Л. М. Давидов}
\address{
 Національний науковий центр ``Харківський фізико-технічний інститут'', Харків, 61108,
вул. Академічна, 1}

\makeukrtitle

\begin{abstract}

Ця робота є продовженням нашої роботи [Condens. Matter Phys., 2020, \textbf{23}, 33602: 1--17]. Вона присвячена моделюванню взаємодії рівноважних і нерівноважних фазових переходів. Моделювання рівноважного фазового переходу базується на модифікованому рівнянні Кана--Хiльярда. Нерівноважний фазовий пере\-хід моделюється другою системою
хімічних реакцій  Шльогля. Розглядається наступаючий фронт, який поєднує ці обидва переходи. На відміну від першої статті, тут ми розглядаємо ефекти пам’яті, тобто ефекти нефікківської дифузії. Отримано розв'язок біжучої хвилі, детально досліджено його залежність від параметрів моделі. Оцінено відносну важливість ефектів пам'яті для різних режимів процесу.

\keywords фазовий перехід, нерівноважний фазовий перехід, рівняння Кана--Хiльярда, реакції
 Шльогля, ефект пам'яті, біжуча хвиля

\end{abstract}

\end{document}